\documentclass[useAMS,usenatbib]{mn2e}
\usepackage{epsfig}
\input psfig.sty

\usepackage{graphicx}% Include figure files
\usepackage{dcolumn}% Align table columns on decimal point
\usepackage{bm}% bold math
\usepackage{epsfig}
\usepackage{graphicx}% Include figure files
\usepackage{dcolumn}% Align table columns on decimal point
\usepackage{bm}% bold math
\usepackage[english]{babel}

\newcommand{\mytilde}{\raise.17ex\hbox{$\scriptstyle\mathtt{\sim}$}}

\title[Robustness of $H_0$ determination]{Robustness of $H_0$ determination at intermediate redshifts}

\author[R. F. L. Holanda, V. C. Busti and G. Pordeus da Silva]{R. F. L. Holanda$^{1,2}$\thanks{E-mail: 
holanda@uepb.edu.br}, V. C. Busti$^{3}$\thanks{E-mail: 
vinicius.busti@iag.usp.br} and G. Pordeus da Silva$^{2}$\thanks{E-mail: 
givalpordeus@hotmail.com} \\ $^{1}$ Departamento de F\'{\i}sica, Universidade Estadual da Para\'{\i}ba, 58429-500 Campina Grande, PB, Brazil \\ $^{2}$ Departamento de F\'{\i}sica, Universidade Federal de Campina Grande, 58429-900 Campina Grande, PB, Brazil \\ $^{3}$  Astrophysics, Cosmology and Gravity Center (ACGC), and  \\ Department of Mathematics and Applied Mathematics, 
University of Cape Town, Rondebosch 7701, Cape Town, South Africa}

\newcommand{\ok}{\Omega_k}

\begin{document}

\date{Accepted 2014 June 4. Received 2014 May 13; in original form 2014 April 16 }

\pagerange{\pageref{firstpage}--\pageref{lastpage}} \pubyear{2014}

\maketitle

\label{firstpage}
\begin{abstract}

\noindent The most recent Hubble constant ($H_0)$ estimates from local methods ($z\ll1$), $H_0=73.8\pm 2.4$ km s$^{-1}$ Mpc$^{-1}$, and the one from high redshifts 
$H_0=67.3\pm 1.2$ km s$^{-1}$ Mpc$^{-1}$,  are discrepant at $2.4 \sigma$ confidence level. Within this context, Lima \& Cunha (LC)  
derived a new determination of $H_0$ using four cosmic probes  at intermediate redshifts ($0.1<z<1.8$) based on the so-called flat $\Lambda$CDM model. 
They obtained $H_0=74.1\pm 2.2$ km s$^{-1}$ Mpc$^{-1}$, in full agreement with local measurements. In this Letter, we explore the robustness of the LC result searching for 
systematic errors and its dependence from the cosmological model used. We find that the $H_0$ value from this joint analysis is very weakly dependent on the underlying cosmological model, 
but the morphology adopted to infer the distance to galaxy clusters changes the result sizeably, being the main source of systematic errors.
Therefore, a better understanding of the cluster morphology is paramount to transform this method into a powerful cross-check for $H_0$.

\end{abstract}


\section{Introduction}

The new controversy in the value of the Hubble constant $H_0$ determined from local and global measurements raised a lot of activity to pin down evidence of new physics or 
unaccounted systematic errors. While a local measurement with Cepheids and Type Ia supernovae (SNe Ia) derived $H_0=73.8 \pm  2.4$ km s$^{-1}$ Mpc$^{-1}$ \citep{riess2011}, 
the Planck Collaboration \citep{planck} determined $H_0=67.3 \pm 1.2$ km s$^{-1}$ Mpc$^{-1}$ within a flat $\Lambda$ cold dark matter ($\Lambda$CDM) model from temperature anisotropies in the cosmic microwave background (CMB).

Many systematic errors may be responsible for the difference. Concerning local measurements, the first rung in the distance ladder is crucial for $H_0$ measurements. Depending on what method is used to calibrate SNe Ia distances, a variety of values is derived for $H_0$. For example, \cite{riess2011} used three distance indicators to calibrate the SNe Ia: a geometric distance to NGC 4258 based on a megamaser measurement; parallax measurements to
Milky Way Cepheids (MWC) and Cepheids observations and
a revised distance to the Large Magellanic Cloud (LMC). Revising the distance to NGC 4258 from \cite{humphreys}, with only this indicator \cite{efs} used \cite{riess2011} data to get $H_0 = 70.6 \pm 3.3$ km s$^{-1}$ Mpc$^{-1}$, while combining the three indicators the value is $H_0 = 72.5 \pm 2.5$ km s$^{-1}$ Mpc$^{-1}$, alleviating the tension. 
On the other hand, calibrating the SNe with the tip of the red giant branch (TGRB) \cite{tammann} obtained $H_0 = 63.7 \pm 2.7$ km s$^{-1}$ Mpc$^{-1}$. This is very intriguing since many local determinations obtained higher values for $H_0$ (see Table 1).

The systematic errors may also come from the CMB analysis. For example, \cite{spergel} claim that the 217 $\times$ 217 GHz detector can be responsible for some part of the tension, where its removal provides $H_0 = 68.0 \pm 1.1$ km s$^{-1}$ Mpc$^{-1}$. Moreover, an inconsistency of the {\it Planck} data with a flat $\Lambda$CDM model was claimed by \cite{sha}, where a lack of power for high and low multipoles may indicate new physics or systematic errors.

On the new physics side, it could be just cosmic variance \citep{bubble,cosmic_variance2}; if we live in a `Hubble bubble' we would infer a higher local value for $H_0$ compared to the global one. This possibility requires a very unlikely size for the void, although it is compatible with observations \citep{keenan}. Other possibilities include extensions of the cosmic concordance model \citep{cde,xcdm} or massive neutrinos \citep{neutrinos}.

One way to avoid local effects is to go to intermediate redshifts. Based on a non-parametric reconstruction of $H(z)$ data, \cite{busti2014} extrapolated the reconstruction to redshift 0 and obtained $H_0 = 64.9 \pm 4.2$ km s$^{-1}$ Mpc$^{-1}$, independent of a cosmological model. Conversely, by adopting a flat $\Lambda$CDM model, \cite{vital2014}\footnote{Following a long series of $H_0$ determinations within a cosmological model, (e.g. \cite{cunha_2007,lima2009,busti2012,holanda_2010}).} used four different probes at intermediate redshifts, namely, angular diameter distances (ADD) from galaxy clusters, 11 ages of old high-redshift galaxies (OHRG), 18 $H(z)$ data points and baryon acoustic oscillations (BAOs) peak. They obtained $H_0=74.1 \pm 2.2$ km s$^{-1}$ Mpc$^{-1}$. While the former value is consistent with {\it Planck}, the latter points to an internal inconsistency inside the $\Lambda$CDM model, since both {\it Planck} and LC values were derived assuming the same model. It is worth mentioning that LC 
used the Bonamente et al. (2006) galaxy clusters sample, modelled by  a non-isothermal spherical double $\beta$ model. However, the standard spherical geometry has been severely questioned, since  {\it Chandra} and {\it XMM-Newton} observations have shown that galaxy clusters exhibit preferably an elliptical X-ray surface brightness (Fox \& Pen 2002; Jing \& Suto 2002; Sereno et al. 2006; Morandi et al. 2010; Limousin et al. 2013). Moreover,  this sample was found to be inconsistent with the so-called cosmic distance duality (DD) relation (Holanda, Lima \& Ribeiro 2010,2011,2012; Meng et al. 2012).

Therefore, the goal of this paper is to analyse the robustness of LC results by performing two kinds of tests. First, we search for systematic errors considering a sample where the 
morphology of galaxy clusters, used to derive their distances, was assumed as isothermal spherical and elliptical.   Also, we test different assumptions for the incubation time $t_{\mathrm{inc}}$ used in the OHRG analysis to see the impact to $H_0$. Secondly, we change the cosmological model to see its dependence on the results, where we consider a curved $\Lambda$CDM model and a flat XCDM model. As we shall see, the $H_0$ value is very weakly dependent on the cosmological model. In contrast, the cluster morphology changes the result sizeably, being the main source of systematic errors, { which means that this source must be controlled in order to claim that this method is a 
powerful cross-check for $H_0$}.

The Letter is organized as follows. In Sec. 2 we introduce the mathematical background and the models adopted in this work. Sec. 3 is devoted to a description of the samples used in the statistical analyses. Sec. 4 presents the results and Sec. 5 closes the paper with the conclusions.

\begin{figure*}
\centering
\includegraphics[width=0.3\textwidth]{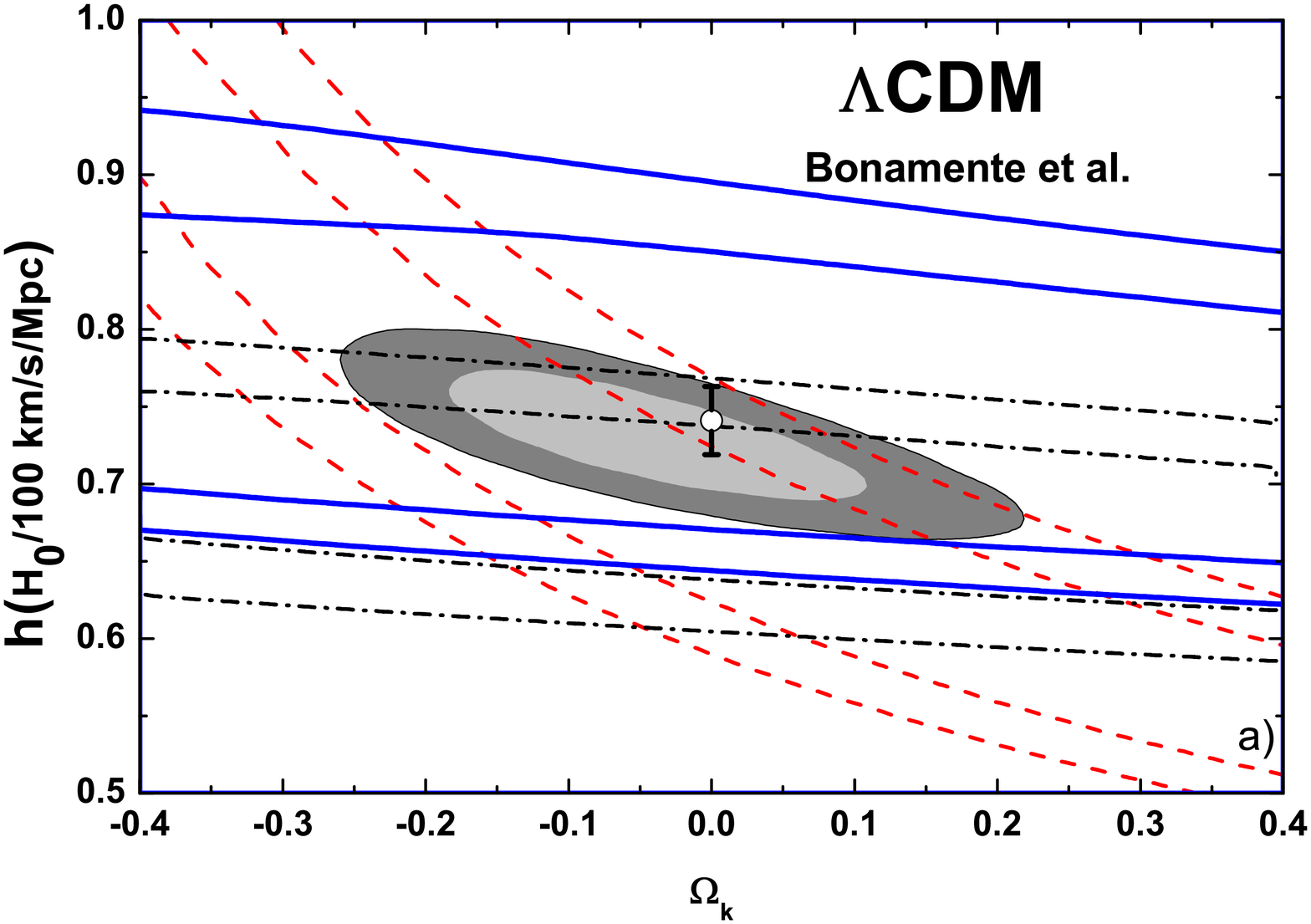}
\includegraphics[width=0.3\textwidth]{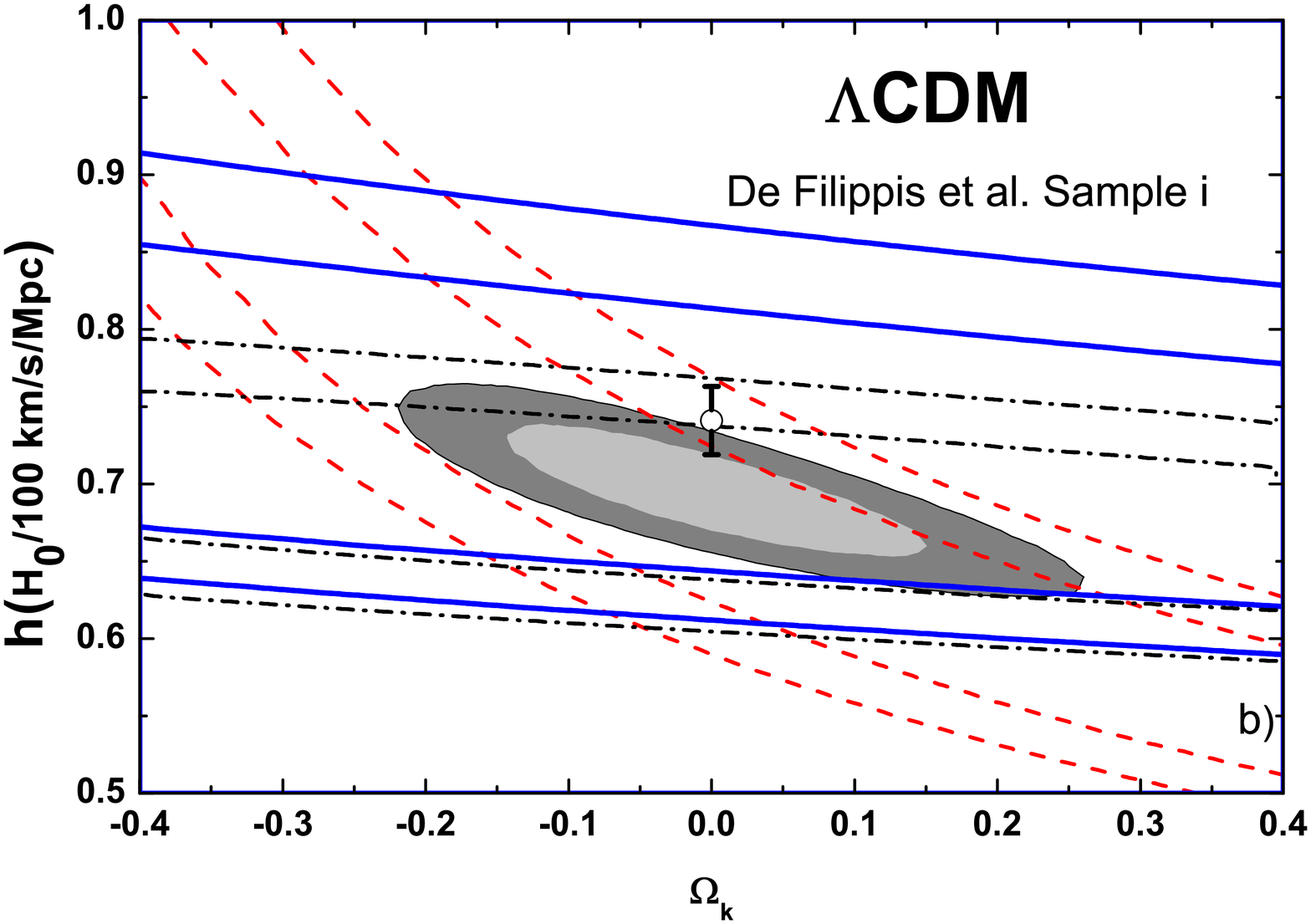}
\includegraphics[width=0.3\textwidth]{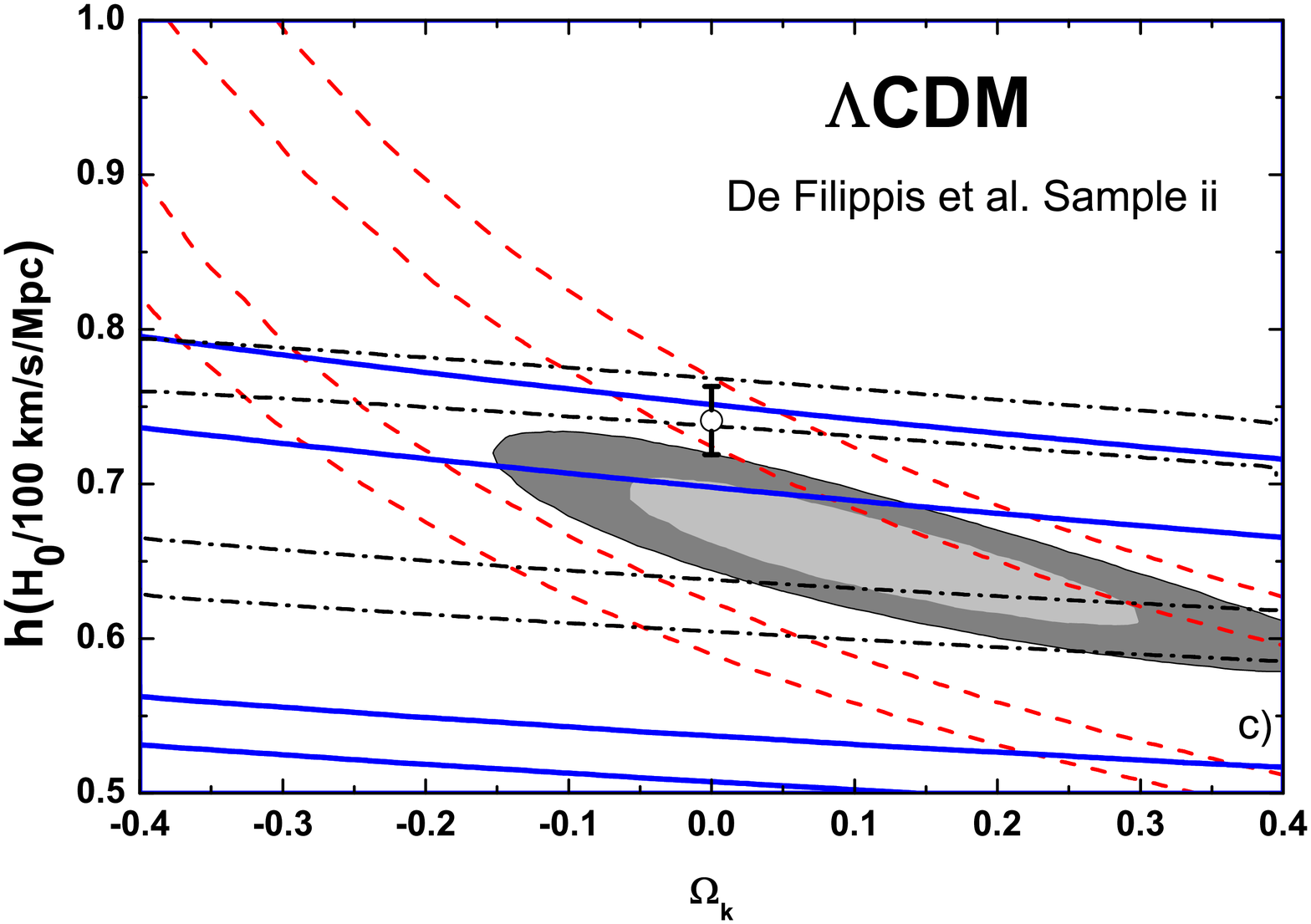}
\caption{$\Lambda$CDM model. In all figures the blue solid, dashed red and dash-dotted black lines correspond to constraints at $68$ and $95$ per cent (c.l.) by using, separately, galaxy clusters,  OHRG+BAO and $H(z)$.(a): the galaxy clusters are from the Bonamente et al. (2006) sample.(b) and (c): the clusters are from De Filippis et al. (2005) samples.  The filled central regions correspond to the joint analysis for each case.  The open circle with its error bar is that one from LC analysis.}
\end{figure*}

\section{Basic Equations and Models}

In this Letter, we consider two cosmological scenarios: (1) the so-called $\Lambda$CDM model, but allowing deviations from flatness, and, (2) the flat XCDM model,  where the universe is flat and driven by CDM$\Omega_{\mathrm{M}}$ plus a dark energy component $\Omega_x$ with constant equation-of-state (EoS) parameter $\omega$. Throughout we use units such that the light speed $c=1$.

By considering the Universe described by a homogeneous and isotropic Friedmann--Lema\^{i}tre--Robertson--Walker (FLRW) geometry, the angular diameter distance $D_{\mathrm{A}}$ is given by 
%\begin{equation}
%\label{line_elem}
%  ds^2 = dt^2 - a^{2}(t) \left(\frac{dr^2}{1-k r^2} + r^2 d\theta^2+
%      r^2{\rm sin}^{2}\theta d \phi^2\right),
%\end{equation}
%where $a(t)$ is the scale factor and $k= 0, \pm 1$ is the curvature
%parameter,  the angular diameter distance ${\cal{D}}_A$ is given by 
\begin{equation}
D_{\mathrm{A}} = \frac{h^{-1}}{(1 +
z)\sqrt{|\ok|}}S_k\left[\sqrt{|\ok|}\int_{0}^{z}\frac{\mathrm{d}z'}{E(z')}\right]
\,\mbox{Mpc}, \label{eq1}
\end{equation}
with $h=H_0/100$ km s$^{-1}$ Mpc$^{-1}$, the function $E(z)={H(z)/H_{0}}$ is the dimensionless Hubble parameter defined by the specific cosmology adopted, $\ok$ is the density curvature parameter and $S_k(x)=\sin{x}, x$, $\sinh{x}$ for $k=+1, 0$, $-1$, respectively.

On the other hand, the age-redshift relation, $t(z)$, is given by $t(z) = \int_{0}^{z}{\mathrm{d}z \over H(z)}$. For the cosmological models adopted in this Letter, $E(z)$ reads $E^2(z)=\frac{H^2(z)}{H^2_0}={\Omega_{\mathrm{M}}(1+z)^3}+{\Omega_x}(1+z)^{3(1+\omega)}+{\Omega_k (1+z)^2}$, 
where $\ok=1-\Omega_x - \Omega_{{\mathrm{M}}}$. We consider two cases: a $\Lambda$CDM model with $\omega=0$ and a flat XCDM model
with $\ok=0$.

\begin{figure*}
\centering
\includegraphics[width=0.3\textwidth]{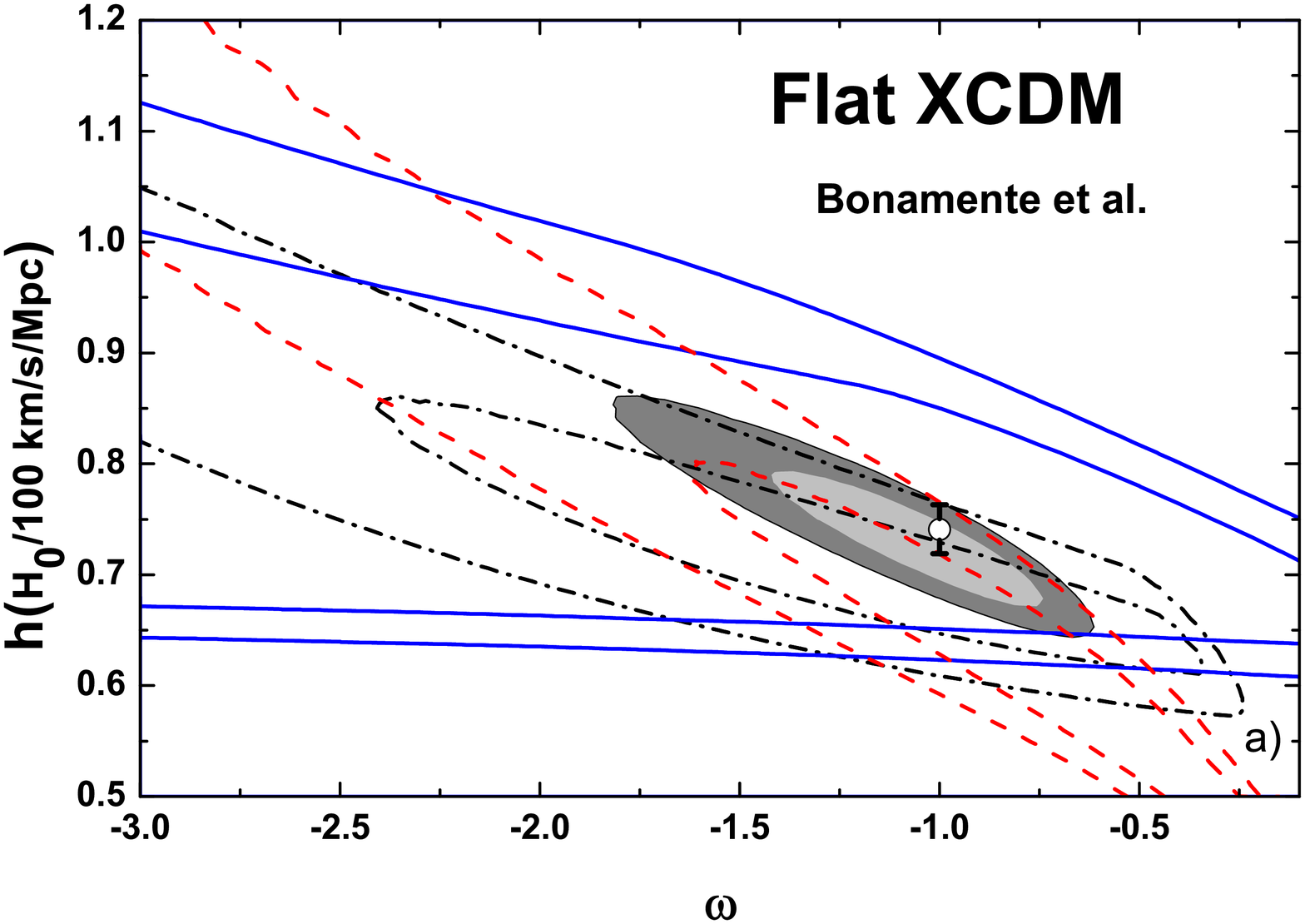}
\includegraphics[width=0.3\textwidth]{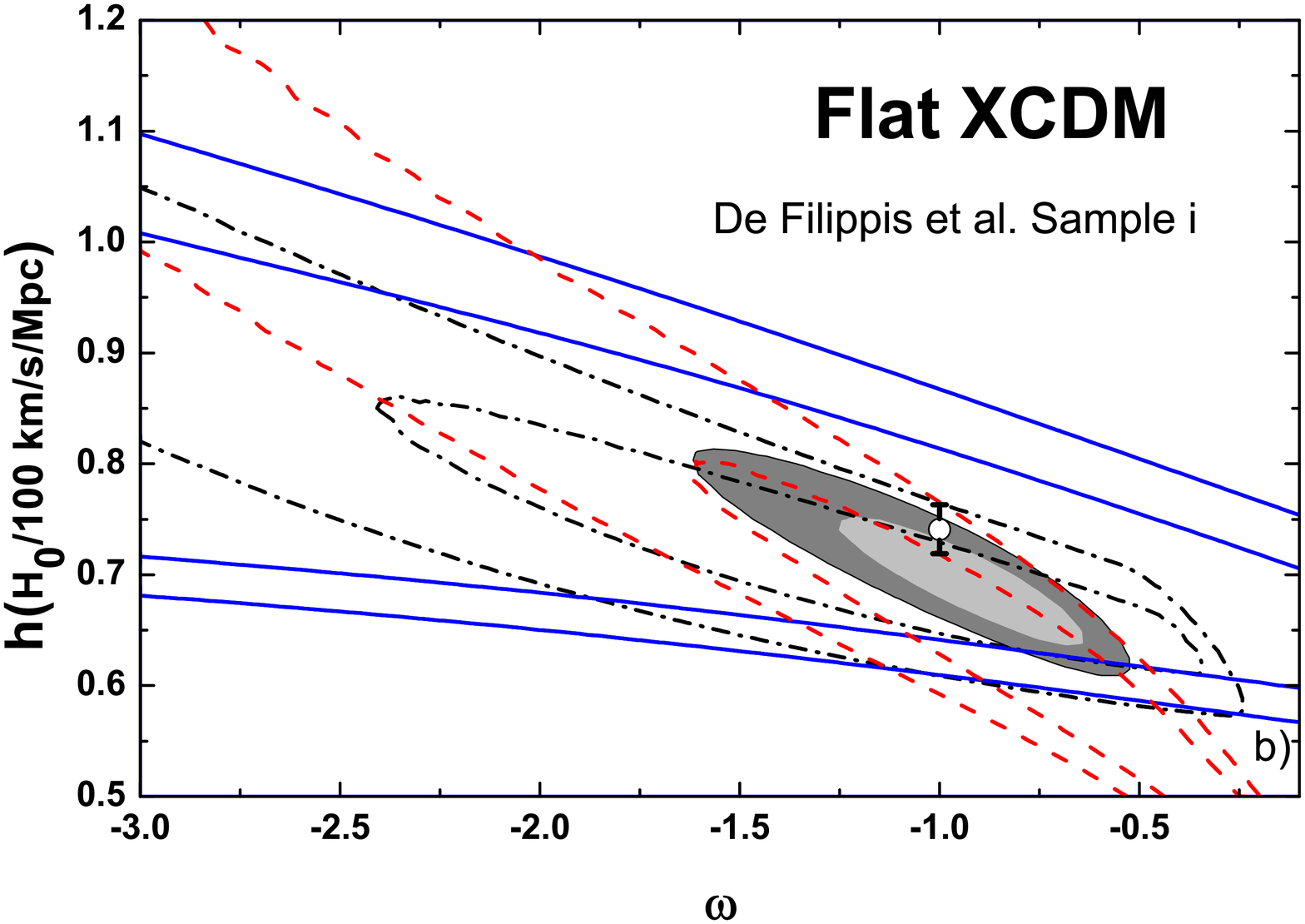}
\includegraphics[width=0.3\textwidth]{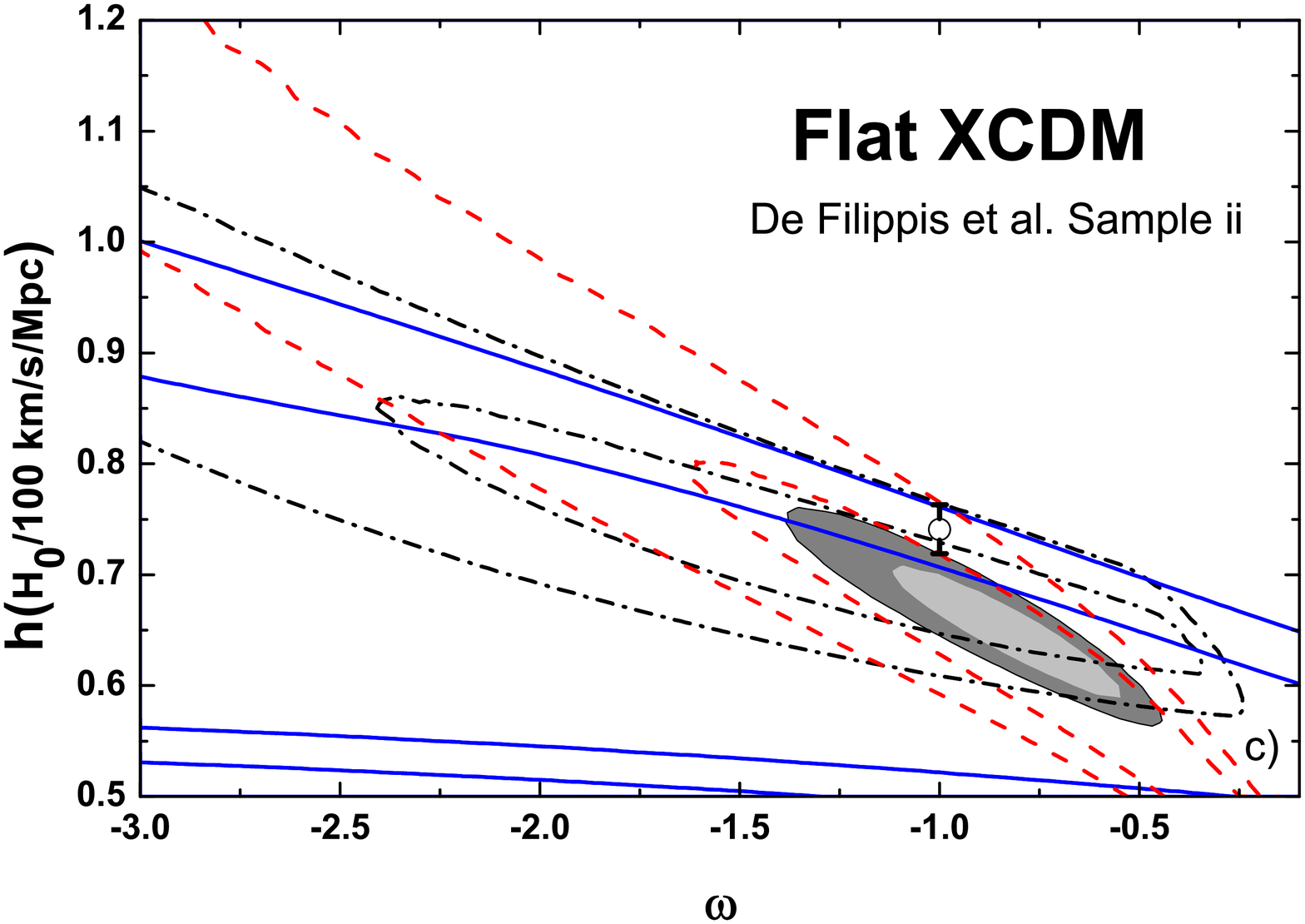}
\caption{The same as Fig. 1 but for a flat XCDM model. 
%In all figures The blue solid, dashed red and dashed-dot black lines correspond to constraints at $68\%$ and $95\%$ (c.l.) by using, separately, galaxy clusters,  OHRG + BAO and $H(z)$. In fig.(a) the galaxy clusters are from the Bonamente et al. (2006) sample. In figs. (b) and (c) the clusters are from De Filippis et al. (2005) samples.  The filled central regions correspond to the joint analysis for each case.  The open circle with its error bar is that one from LC analysis.
}
\end{figure*}
\section{Samples}

In order to have a reliable comparative study with the LC results, the only difference between their and our  cosmological probes it is the addition of the De Filippis et al. sample (2005). 
Summarizing, in this Letter we use the following.

(a) Three samples of ADD from galaxy clusters obtained from their Sunyaev--Zel'dovich and X-ray observations (the so-called ESZ/X-ray technique). The first one, used in LC, composed 
of 38 ADD in redshift range $0.1<z<0.89$ compiled by Bonamente et al. (2006) where the cluster plasma and dark matter distributions 
were analysed assuming  a non-isothermal spherical double $\beta$ model. { This model generalizes the single $\beta$ model proposed by Cavaliere \& Fusco-Fermiano (1978). 
Summarizing, the cluster plasma and dark matter distributions were analysed assuming a hydrostatic equilibrium model and spherical symmetry, accounting for radial variations in
density, temperature and including the possible presence of cooling flow. The  new samples used here are formed by 18 galaxy clusters from De Filippis {{et al.}} (2005)  
in the redshift range $0.142<z<0.79$. These authors re-analysed archival X-ray data of the {\it XMM-Newton} and {\it Chandra} satellites of two samples (Mason et al. 2001; Reese et al. 2002) 
for which combined X-ray and SZE analysis have already been reported. In the reanalysis were used two models to describe exactly the same clusters: the isothermal elliptical and 
spherical  $\beta$ models,  providing two ADD samples, named from now on, samples i and ii, respectively. }

{ It is important to comment that, in general, different cluster gas profiles do not affect the inferred central surface brightness ($S_{x0}$) or central 
Sunyaev--Zel'dovich decrement ($\Delta T_0$), but give different $\theta_c$ (the core radius).  De Filippis et al. (2005) found, for instance,  
$\theta_{ell} = \frac{2e_{proj}}{1+e_{proj}}\theta_{circ}$ (in first approximation), where  $\theta_{ell}$ and $\theta_{circ}$ are the core radius obtained by using an 
isothermal elliptical $\beta$ model and an isothermal spherical $\beta$ model, respectively, and $e_{proj}$ is the axial ratio of the major to the minor axes of the projected isophotes. 
Since $D_{\mathrm{A}}(z)\propto 1/\theta_{\mathrm{c}}$, different core radius affect the ESZ/X-ray distances and, consequently, the $H_0$ estimates (see fig. 1 in De Filippis et al. 2005). 
Thus, for these single $\beta$ models, $D_{\mathrm{A}}$ obtained by the spherical model is overestimated compared with the elliptical one.}

(b) 18 Hubble parameter versus redshift data points, $H(z)$,  from cosmic chronometers and BAOs in redshift range $0.1 < z < 1.8$ (Simon et al. 2005, Gaztanaga et al. 2009, 
Stern et al. 2010). 

(c) The inferred ages of 11 OHRG $(0.62 < z < 1.70)$. These 11 data points are subsamples from Ferreras et al. (2009) and Longhetti et al. (2007) 
catalogues. As  argued by LC, the selected data set provides  accurate and restrictive galaxy ages (see fig. 1 in their paper).

(d) The BAOs peak at $z=0.35$. As it is largely known,  the relevant distance measure is the  dilation scale  that can be modelled as the cube root 
of the radial dilation times the square of the transverse dilation, at the typical redshift of the galaxy sample, $z=0.35$ (Eisenstein et al. 2005):
\begin{equation}
  D_V(z)=[D_{\mathrm{A}}(z)^2 z/H(z)]^{1/3}\ .
\end{equation}
However, the  BAO quantity that we use is the $H_0 $ independent BAO datum given by
\begin{equation}
A(0.35)=D_V(0.35)\frac{\sqrt{\Omega_{\mathrm{M}} H_0^2}}{0.35}=0.469\pm 0.017.
\end{equation}
 
\section{analyses and results}

We perform the
$\chi^{2}$ statistics  combining the
four tests discussed above such as (LC)

\begin{eqnarray}
 \chi^2(z|\mathbf{p})  = & \sum_i { ({\cal{D}}_{\mathrm{A}}(z_i; \mathbf{p})-
{\cal{D}}_{\mathrm{A obs,}i})^2 \over \sigma_{{\cal{D}}_{\mathrm{A obs,}i}}^2 } + \sum_j { (t(z_j;
\mathbf{p})-t_{\mathrm{inc}}- t_{\mathrm{obs,}j})^2 \over
\sigma_{t_{\mathrm{obs,}j}}^2+\sigma_{t_{\mathrm{inc}}}^2} &
\nonumber
\\ &+ \sum_k { (H(z_k; \mathbf{p})-H_{\mathrm{obs,}k})^2 \over
\sigma_{H_{\mathrm{obs,}k}}^2} + {(A(\mathbf{p})-0.469)^2 \over 0.017^2}  .
\end{eqnarray}

The quantities with subscript 'obs' are the observational quantities, $\sigma_{{\cal{D}}_{\mathrm{Aobs,}i}}$ is the uncertainty in the
individual distance, $\sigma_{t_{\mathrm{inc}}}$ is the incubation time
error. For the galaxy cluster samples, the common statistical contributions are SZE point sources $\pm 8\%$, X-ray background $\pm2\%$, galactic NH $< \pm1\%$, $\pm15\%$ for cluster asphericity, $\pm8\%$ kinetic SZ and for CMB anisotropy  $<\pm2\%$. The term $t_{inc}$  is the incubation time,  defined by the amount of time interval from the beginning of structure formation process in the Universe until the formation time  of the object itself. The complete set of parameters is given by $\mathbf{p}$. Following LC, we have considered initially $t_{\mathrm{inc}}=0.8 \pm 0.4$ Gyr. 

\begin{figure*}
\centering
\includegraphics[width=0.35\textwidth]{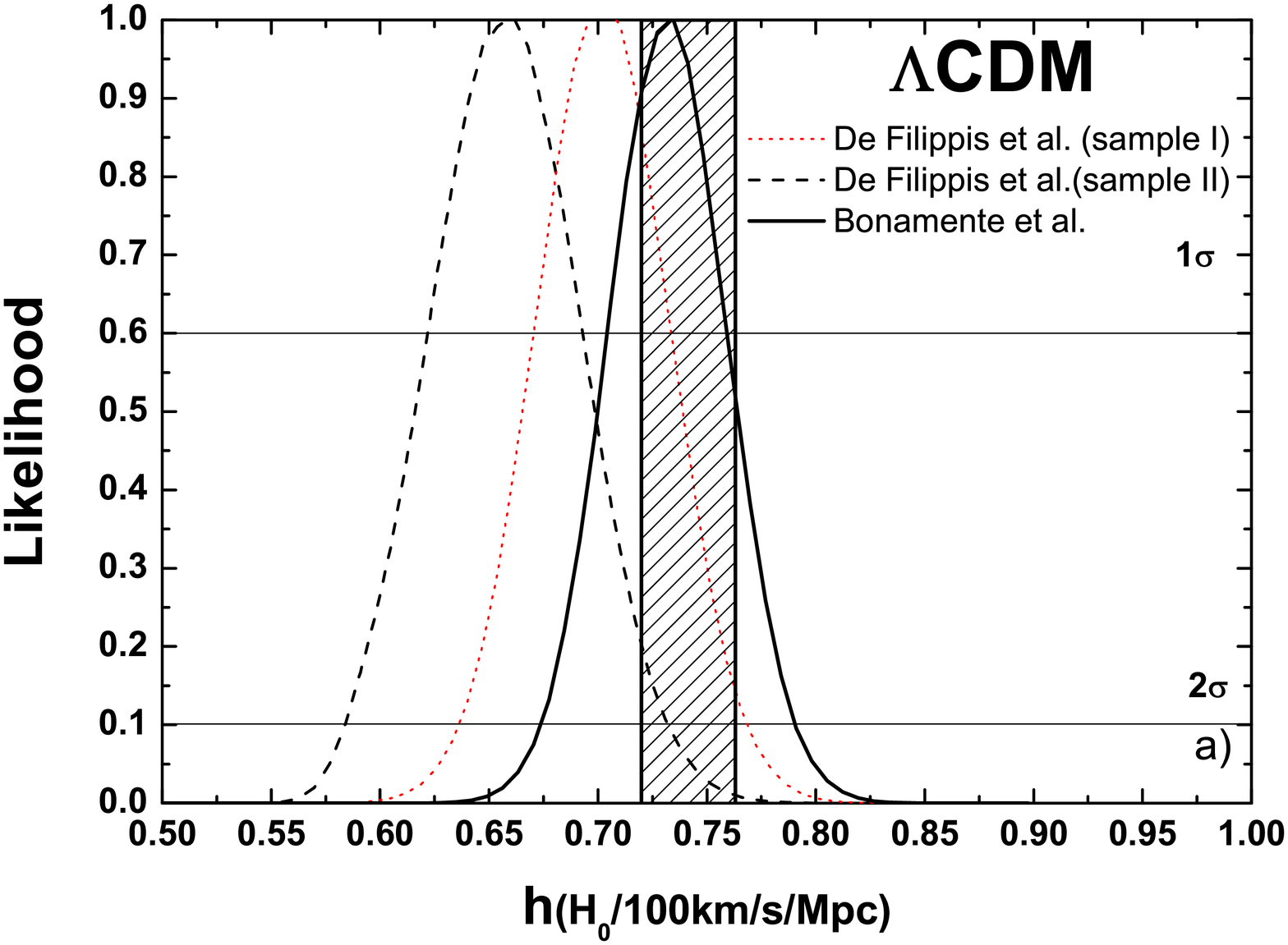}
\includegraphics[width=0.35\textwidth]{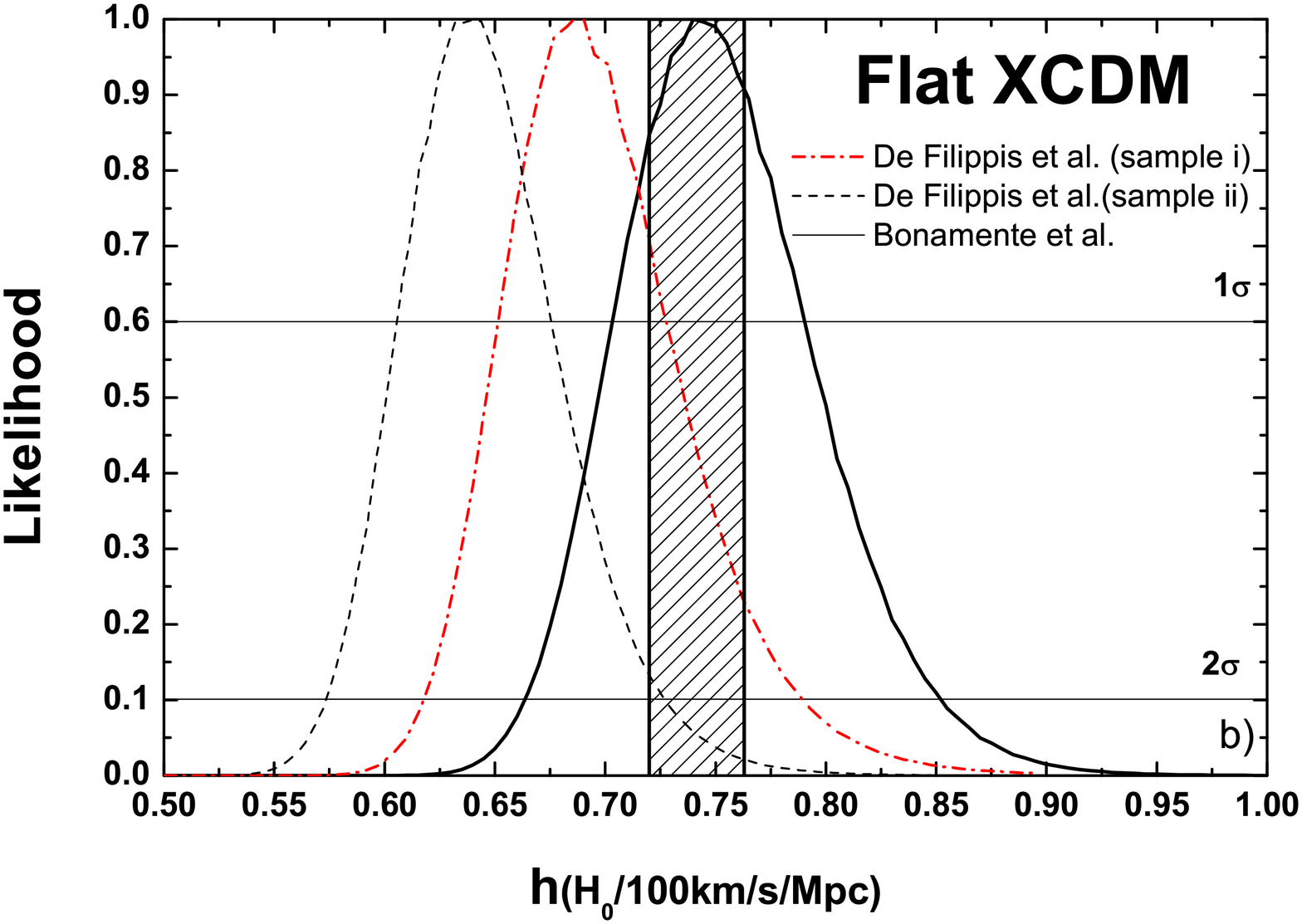}
\caption{Likelihood  of the $h$ parameter. (a) and (b): correspond to $\Lambda$CDM and flat XCDM models, respectively, by including only statistical errors. The black solid, blue dotted and red dash-dotted lines are from analyses by using Bonamente et al. and De Filippis et al. samples i and ii, respectively. The shaded region is the $1\sigma$ interval of the LC analysis. 
}
\end{figure*}

\begin{figure*}
\centering
\includegraphics[width=0.35\textwidth]{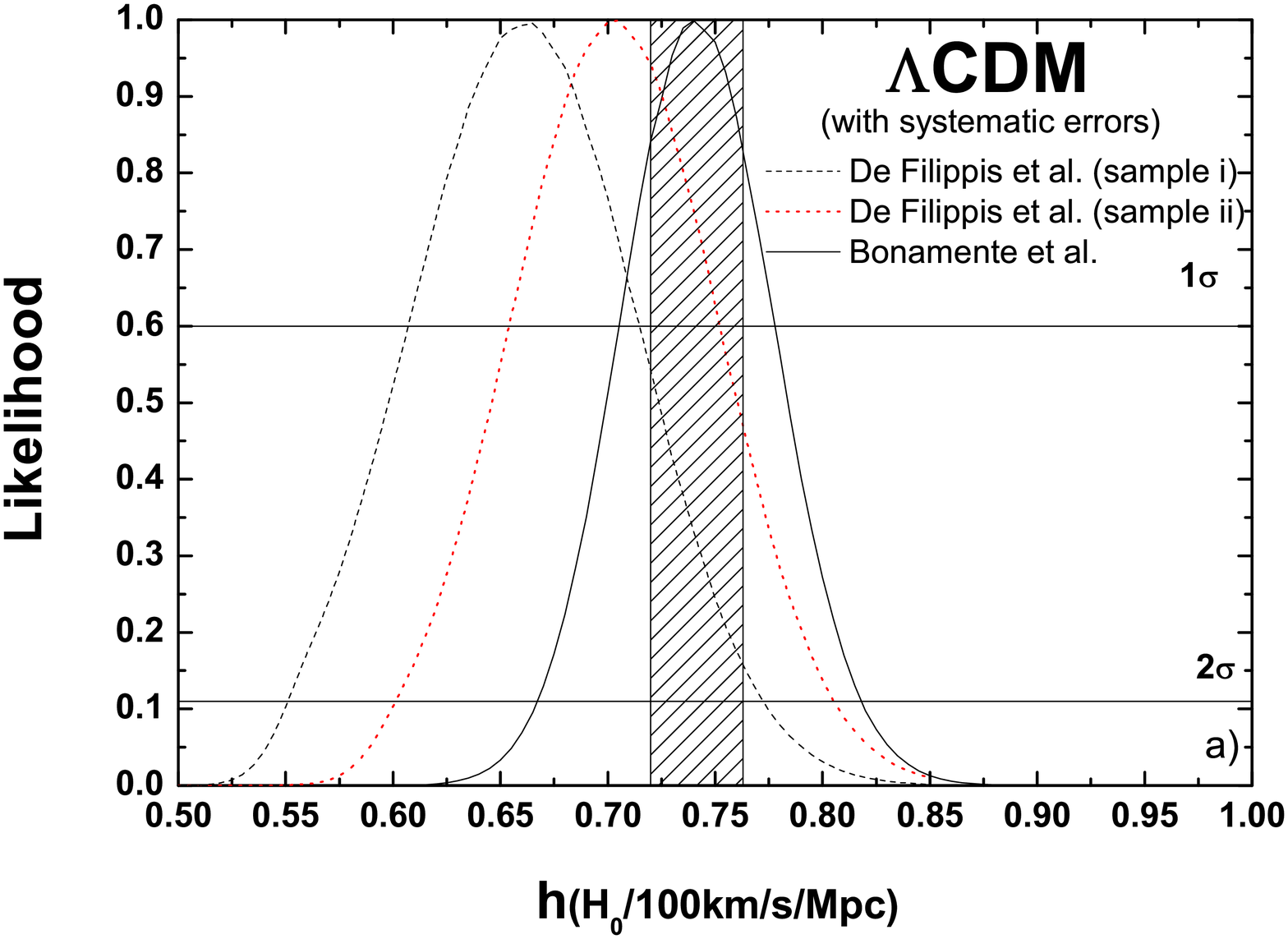}
\includegraphics[width=0.35\textwidth]{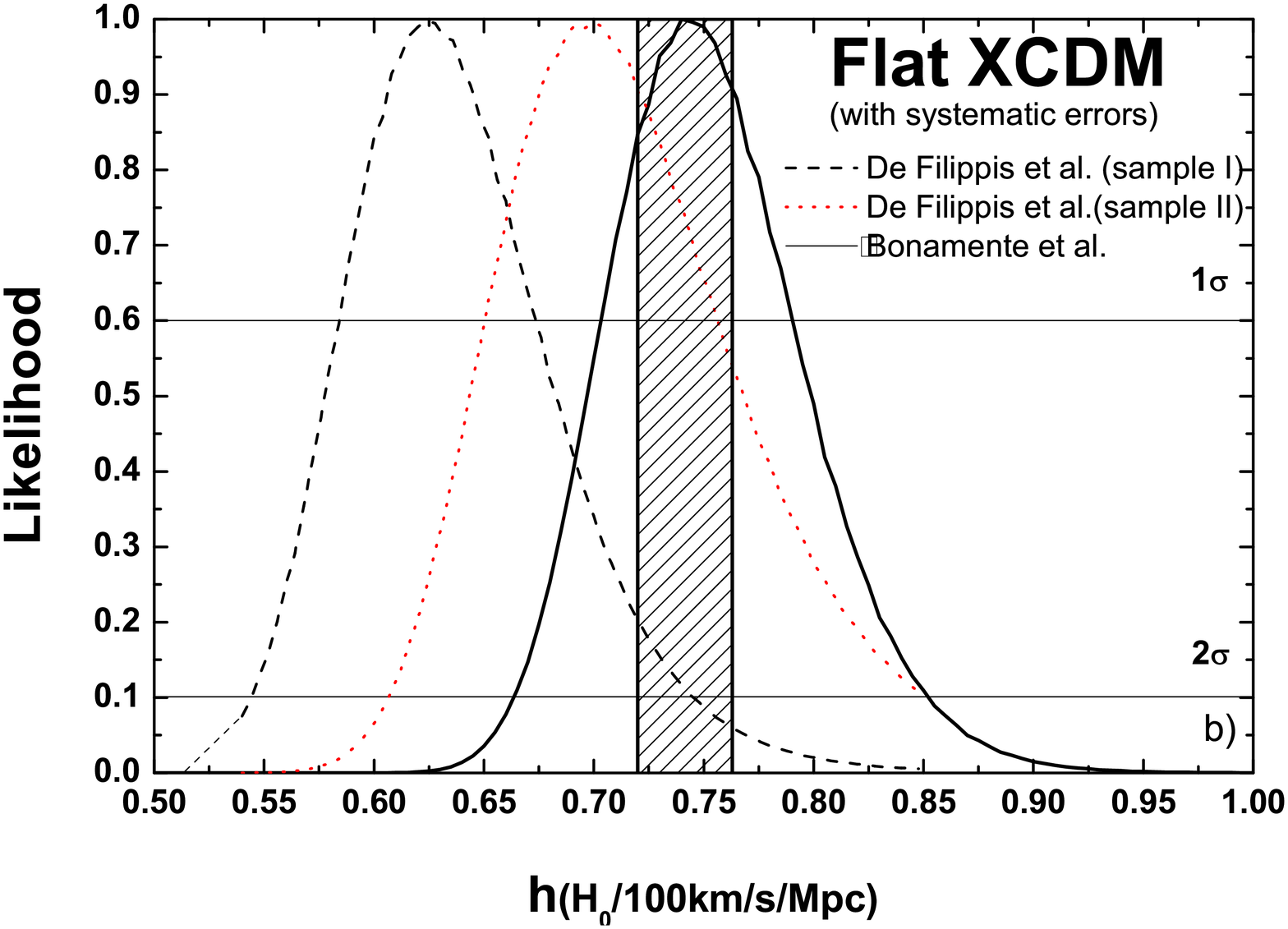}
\caption{Likelihood  of the $h$ parameter. (a) and (b): correspond to $\Lambda$CDM and flat XCDM models, respectively, by adding statistical and systematic errors in 
quadrature. The correspondent lines are the same as in Fig.3.
}
\end{figure*}
\subsection{$\Lambda$CDM} 

In  Figs 1(a)-(c) we display the $(h, \Omega_k)$ plane. Here and in the flat XCDM analyses, the red-dashed and black dash-dotted lines correspond to 1$\sigma$ and 2$\sigma$ 
limits obtained by using  OHRG+BAO and $H(z)$, respectively. The blue solid lines correspond to limits from galaxy clusters samples: Bonamente et al. and De Filippis et al. 
samples i and ii, respectively. For $\Lambda$CDM model, there is a degeneracy between $\Omega_{k}$ and $h$ for all cosmological probes, and, therefore, the possible values for $h$ 
are weakly constrained by data separately. The filled central regions correspond to the joint analysis. The open circle with its error bar is that one from LC analysis. 

From the joint analysis by using the galaxy clusters (Bonamente et al. sample)+OHRG+BAO+$H(z)$  we obtain in figure (1a) for two free parameters: $h=0.74^{+0.04}_{-0.04}$,  
$\ok=-0.044^{+0.14}_{-0.15}$ and $\chi^2_{red}=0.98$ at $68.3$\% (c.l.) . This $h$ estimate  is in full agreement with the LC value, being the constraints on $h$  independent of a 
flat universe assumption. However, by using the other galaxy cluster samples in the joint analysis, we obtain in the figures (1b) and (1c): $h=0.70^{+0.04}_{-0.04}$,  
$\ok=0.012^{+0.14}_{-0.14}$ and $\chi^2_{red}=0.96$ and $h=0.65^{+0.06}_{-0.06}$,  $\ok=0.14^{+0.15}_{-0.19}$ and $\chi^2_{red}=0.94$, respectively.  

In fig. 3(a) we display the likelihood function of the $h$ parameter. To obtain this we marginalized over $\Omega_{\mathrm{M}}$ and $\Omega_{\Lambda}$ parameters. The horizontal lines 
are cuts in the probability regions of $68.3$ and $95.4$ per cent. For this case we obtain, at 1$\sigma$,  $h=0.74^{+0.035}_{-0.030}$, $0.70^{+0.035}_{-0.037}$ 
and $0.65^{+0.042}_{-0.042}$ for Bonamente et al. and De Filippis et al. samples i and ii, respectively. The shaded region corresponds to 1$\sigma$ interval derived by LC.  
We also performed the analysis by using the De Filippis et al. samples in a flat $\Lambda$CDM, obtaining $h=0.705^{+0.020}_{-0.018}$ and $0.65^{+0.020}_{-0.021}$ for 
samples i and ii, respectively, incompatible at least in 1$\sigma$ with the one obtained by LC ($0.719<h<0.763$).

\subsection{Flat XCDM}

In  Figs  2(a)-(c) we display the $(h, \omega)$ plane. There is a strong dependence between $h$ and $\omega$ for all cosmological probes. The filled central regions 
correspond to the joint analyses. Again, the open circle with its error bar is that one from LC analysis. 

From the joint analysis by using the galaxy clusters+OHRG+BAO+$H(z)$  we obtain in fig. 2(a) for two free parameters $h=0.72^{+0.06}_{-0.06}$,  $\omega=-1.1^{+0.50}_{-0.45}$ 
and $\chi^2_{red}=0.97$ at $68.3$\% (c.l.). This $h$ estimate  is in agreement with the LC value, being, therefore, the constraints on $h$  independent of $\omega=-1$ assumption. 
However, again,  from the joint analyses in the figures 2(b) and (c), we obtain $h=0.68^{+0.07}_{-0.05}$,  $\omega=-0.73^{+0.45}_{-0.5}$ and $\chi^2_{red}=0.95$ and $h=0.64^{+0.06}_{-0.06}$,  $\omega=-0.73^{+0.45}_{-0.5}$  and $\chi^2_{red}=0.94$, respectively.  As one may see, it is strongly dependent on the model used to describe the galaxy clusters. 

In Fig. 3(b) we display the likelihood function of the $h$ parameter. To obtain this graph we have marginalized over $\Omega_{\mathrm{M}}$ and $\omega$ parameters. The horizontal lines are cuts 
in the probability regions of $68.3$ and $95.4$ per cent. For this case we obtain, at 1$\sigma$, $h=0.74^{+0.050}_{-0.050}$, $0.69^{+0.04}_{-0.04}$ and $0.64^{+0.041}_{-0.038}$ for 
Bonamente et al. and De Filippis et al. samples i and ii, respectively. The shaded region corresponds to the 1$\sigma$ interval derived by LC for a flat $\Lambda$CDM model. 

As one may see, regardless of cosmological model used, the $H_0$ values obtained here  by using the Bonamente et al. sample  are in agreement with that one from Lima \& Cunha (2014)  
performed in a flat $\Lambda$CDM model. Moreover, the $H_0$ estimates from  De Filippis et al. samples also are independent of the underlying cosmological model used. 
However, the $h$ estimates in all cases are dependent on the model used to describe the galaxy clusters with a { mild} tension between Bonamente et al. sample and De Filippis et al. 
sample ii. 

\subsection{Systematic errors}

Besides the cluster morphology, there are other sources of systematic errors which can affect the constraints.
Therefore, we redid all analyses including the systematic errors in quadrature (see Figs 4a and b) used in LC analysis, they are { $8$ per cent on $H(z)$ and $15$ per cent on OHRG ages, 
moreover, from SZE/X-ray technique we have  SZ
calibration $\pm 8$ per cent, X-ray flux calibration $\pm 5$ per cent, radio haloes
$+3$ per cent and X-ray temperature calibration $\pm 7.5$ per cent.  As a matter of
fact, one may show that typical systematic errors amount for nearly
$13$ per cent (details can be found in Bonamente et al. 2006)}. We obtain at 1$\sigma$ for Bonamente et al. and De Filippis et al. samples i and ii, respectively:  
(1) $\Lambda$CDM: $h=0.74^{+0.043}_{-0.038}$, $0.705^{+0.045}_{-0.045}$ and $0.64^{+0.062}_{-0.070}$; (2) flat XCDM: $h=0.74^{+0.051}_{-0.048}$, $0.69^{+0.060}_{-0.050}$ 
and $0.63^{+0.060}_{-0.058}$.  We also performed the analysis by using the De Filippis et al. samples in a flat $\Lambda$CDM, obtaining $h=0.70^{+0.03}_{-0.028}$ 
and $0.65^{+0.03}_{-0.03}$ for samples i and ii, respectively. Thus, for all cases, the tension between LC analysis and the De Filippis et al. sample ii is still at 
least 1$\sigma$, with the last one  preferring low $H_0$ value in agreement with Busti et al. 2014. On the other hand, the results by using De Filippis et al. sample i  are  in full 
agreement with that one from 9-year \textit{Wilkinson Microwave Anisotropy Probe} (\textit{WMAP}9; see Table 1). { It is important to comment that numerical simulations (Sulkanen 1999) showed that a spherical model fit to triaxial X-ray 
and SZE clusters should provide an unbiased estimate of $H_0$  when a large ensemble
of clusters are used (since elongated clusters give $H_0$ underestimated), which is not the case at the moment.} Moreover, we also explored a possible dependence on $h$ estimates 
arising from the chosen incubation time. Thus, we changed the  incubation time to $1.2\pm 0.6 $ and $0.4\pm 0.2$ Gyr and redid our analyses. The influence found was negligible. 

\begin{table}
\caption{Constraints on $h$ for different methods  (statistical plus systematic errors). Comb1 stands for {{SZE/X-ray+age+$H(z)$+BAO (Bonamente sample)}}, Comb2 for the combination 
with the De Filippis sample i and Comb3 for the De Filippis sample ii.}
{\begin{tabular} {c||c||c}
Reference & Method & $h$ ($1\sigma$) \\
\hline \hline 
Chen \& Ratra 2011& Median Statistics & $0.680\pm 0.028$  \\
Hinshaw {\it{et al.}} 2013& \textit{WMAP}9 &$0.700\pm 0.022$  \\
Freedman {\it{et al.}} 2012& SNe Ia/Cepheid &$0.743\pm 0.026$ \\
Ade {\it{et al.}} (2013)& \textit{Planck} &$0.673\pm 0.012$   \\
Busti {\it{et al.}} 2014& $H(z)$ &$0.649\pm 0.042$   \\
{{LC (flat $\Lambda$CDM)}}& Comb1 &$0.741\pm 0.022$   \\
 This Letter (flat $\Lambda$CDM)& Comb2 &$0.70\pm0.03$  \\
 This Letter (flat $\Lambda$CDM)& Comb3 &$0.65\pm0.03$  \\
 This Letter ($\Lambda$CDM)& Comb1 &$0.74^{+0.043}_{-0.038}$ \\
 This Letter ($\Lambda$CDM)& Comb2 &$0.705^{+0.045}_{-0.045}$ \\
 This Letter ($\Lambda$CDM)& Comb3 &$0.64^{+0.062}_{-0.070}$ \\
  This Letter (flat XCDM)& Comb1 &$0.74^{+0.051}_{-0.048}$ \\
 This Letter (flat XCDM)& Comb2 &$0.70^{+0.03}_{-0.028}$ \\
 This Letter (flat XCDM)& Comb3 &$0.65^{+0.03}_{-0.03}$ \\
\hline
\end{tabular}} \label{ta2}
\end{table}
\section{Conclusions}

{ In this work we have discussed the robustness of determination of the Hubble
 constant by using the following cosmic probes at  intermediate redshifts: (i) angular diameter distances for galaxy clusters, (ii) the inferred ages of OHRG, 
 (iii) measurements of the Hubble parameter and (iv) the BAO signature. In the angular diameter distances we consider  three samples of galaxy clusters from 
 Bonamente et al. (2006) and De Filippis et al. (2005), which use different assumptions on the galaxy clusters properties.  As emphasized by LC, the combination of these four independent
phenomena at intermediate redshifts is independent of any calibrator usually adopted in the determinations of the distance scale.

 From our results, we conclude that the $H_0$ estimates present a negligible dependence on dark energy models and the incubation time of the OHRG analysis. However, even taking 
 into account statistical and systematic errors, the galaxy clusters data proved to be an important source of systematic errors (see Table 1), making this technique at the moment 
 unable to discriminate between the local value obtained by Riess et al. (2011) and the global value determined by \textit{Planck} as claimed by LC.}
%----------------------------------------------------------%
\section*{Acknowledgements}
%----------------------------------------------------------%
RFLH is supported by INCT-A and CNPq (no. 478524/2013-7). VCB is supported by CNPq - Brazil through a fellowship within the program Science without Borders, and GPdS is supported by CAPES.

\end{document}